\documentclass[twocolumn,prb,superscriptaddress,showpacs,floatfix]{revtex4}

\usepackage{graphicx}

\begin{document}

\title{Magnetic structure of Cd-doped CeCoIn$_5$}

\author{M. Nicklas}\email{nicklas@cpfs.mpg.de}
\affiliation {Max Planck Institute for Chemical Physics of Solids,
N{\"o}thnitzer Str. 40, 01187 Dresden, Germany}

\author{O. Stockert}
\affiliation {Max Planck Institute for Chemical Physics of Solids,
N{\"o}thnitzer Str. 40, 01187 Dresden, Germany}
\author{Tuson Park}
\affiliation {Los Alamos National Laboratory, Los Alamos, NM 87545,
USA}
\author{K. Habicht}
\affiliation {Hahn-Meitner-Institut, Glienicker Str. 100, 14109
Berlin, Germany}
\author{K. Kiefer}
\affiliation {Hahn-Meitner-Institut, Glienicker Str. 100, 14109
Berlin, Germany}
\author{L. D. Pham}
\affiliation {University of California, Davis, CA 95616,
USA}
\author{J. D. Thompson} \affiliation {Los Alamos National
Laboratory, Los Alamos, NM 87545, USA}
\author{Z. Fisk}
\affiliation {University of California, Irvine, CA 92697, USA}
\author{F. Steglich}
\affiliation {Max Planck Institute for Chemical Physics of Solids,
N{\"o}thnitzer Str. 40, 01187 Dresden, Germany}

\date{\today}

\begin{abstract}
The heavy fermion superconductor CeCoIn$_5$ is believed to be close
to a magnetic instability, but no static magnetic order has been
found. Cadmium doping on the In-site shifts the balance between
superconductivity and antiferromagnetism to the latter with an
extended concentration range where both types of order coexist at
low temperatures. We investigated the magnetic structure of
nominally 10\% Cd-doped CeCoIn$_5$, being antiferromagnetically
ordered below $T_{\rm N}\approx 3$~K and superconducting below
$T_c\approx1.3$~K, by elastic neutron scattering. Magnetic intensity
was observed only at the ordering wave vector $Q_{\rm AF} =
(\frac{1}{2},\frac{1}{2},\frac{1}{2})$ commensurate with the crystal
lattice. Upon entering the superconducting state the magnetic
intensity seems to change only little. The commensurate magnetic
ordering in CeCo(In$_{1-x}$Cd$_x$)$_5$ is in contrast to the
incommensurate antiferromagnetic ordering observed in the closely
related compound CeRhIn$_5$. Our results give new insights in the
interplay between superconductivity and magnetism in the family of
Ce$T$In$_5$ ($T= {\rm Co}$, Rh, and Ir) based compounds.

\end{abstract}


\maketitle

In conventional superconductors only small amounts of magnetic
impurities destroy the superconducting state, while in the
heavy-fermion systems, like CeCu$_2$Si$_2$,\cite{Steglich79} the
presence of a dense lattice of magnetic rare-earth atoms is needed
to generate unconventional superconductivity (SC). The discovery of
SC and antiferromagnetism in the family of Ce$T$In$_5$ ($T= {\rm
Co}$, Rh, or Ir) compounds, forming in the tetragonal HoCoGa$_5$
structure, with CeCoIn$_5$ \cite{Petrovic01b} and CeIrIn$_5$
\cite{Petrovic01a} displaying a superconducting ground state and
CeRhIn$_5$ \cite{Hegger00} being antiferromagnetically ordered,
offers an ideal opportunity to study the peculiar interplay of these
two ground states. Though no sign of static magnetic order has been
found in CeCoIn$_5$, pronounced non-Fermi-liquid (NFL) behavior in
the normal state in thermodynamic and transport properties is
observed at low
temperatures.\cite{Petrovic01b,Bianchi03,Paglione03,Kohori01}
Commonly, NFL behavior occurs in the close proximity to a quantum
critical point (QCP). The search for a magnetic phase in CeCoIn$_5$,
which could give rise to such a QCP by a continuous suppression of
the transition temperature via, e.g., applying external pressure or
chemical doping has not been successful until recently. Studies of
Cd doping on the In-site in CeCoIn$_5$ revealed that only a few
percent of Cd doping indeed lead to the development of
antiferromagnetic (AF) order.\cite{Pham06} Also, reports on the
existence of field-induced magnetism in the Abrikosov vortex state
in CeCoIn$_5$ underline its proximity to
magnetism.\cite{Mitrovic06,Young07}

\begin{figure}[b]
\centering
\includegraphics[angle=0,width=70mm,clip]{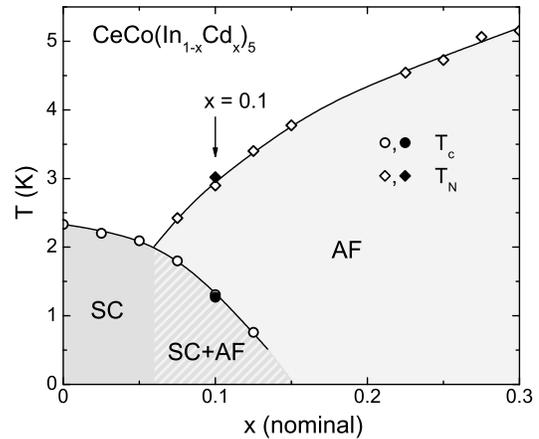}
\caption{Doping - temperature ($x-T$) phase diagram of
CeCo(In$_{1-x}$Cd$_x$)$_5$, where $x$ is the nominal Cd
concentration.  Diamonds indicate the N\'{e}el temperature, $T_{\rm
N}$, and circles the superconducting transition temperature, $T_c$,
determined by specific heat measurements. The arrow indicates the
concentration investigated in this work, with transition
temperatures as indicated by the solid symbols. Open symbols
represent data taken from Ref.~8.\label{PhD}}
\end{figure}

Cd-doping continuously suppresses the superconducting transition
temperature of CeCoIn$_5$ ($T_c=2.3$ K), as shown in the phase
diagram depicted in Fig.~\ref{PhD}. For a nominal Cd-concentration
in excess of 7.5\%, AF order has been observed, coexisting with SC
at low temperatures.\cite{Pham06} With further increasing
Cd-concentration the N\'{e}el temperature, $T_{\rm N}$, increases
monotonically and no SC is found above 12.5\% Cd anymore. In this
report we present neutron scattering data on a
CeCo(In$_{1-x}$Cd$_x$)$_5$ sample with $x=0.1$, where $x$ represents
the nominal concentration, situated in the concentration range where
SC and antiferromagnetism coexist at low temperatures.

\begin{figure}[t]
\centering
\includegraphics[angle=0,width=75mm,clip]{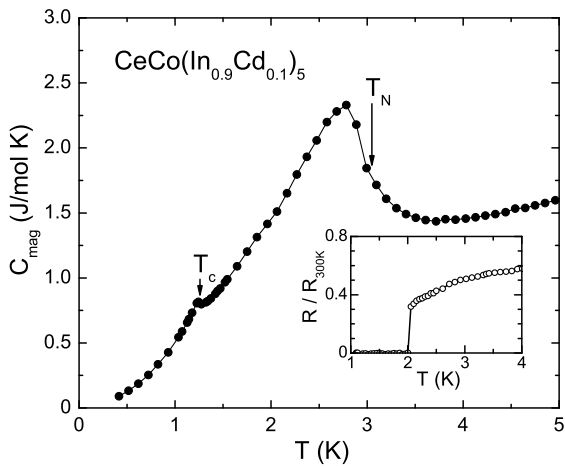}
\caption{Temperature dependence of the magnetic contribution to the
specific heat, $C_{\rm mag}$, for CeCo(In$_{0.9}$Cd$_{0.1}$)$_5$. To
obtain $C_{\rm mag}$, the contribution of the non-magnetic reference
compound LaCoIn$_5$ \cite{Hundley} was subtracted from the specific
heat of CeCo(In$_{0.9}$Cd$_{0.1}$)$_5$. The N\'{e}el temperature,
$T_{\rm N}$, and the superconducting transition temperature, $T_c$,
are indicated by arrows. Inset shows the electrical resistivity
measured on the same sample.\label{Cp_rho}}
\end{figure}

Single crystals of CeCo(In$_{0.9}$Cd$_{0.1}$)$_5$ were grown using a
standard In-flux technique with a nominal concentration of 10\% Cd
in the indium flux. X-ray diffraction confirmed that the samples
crystallize in the tetragonal HoCoGa$_5$ type of crystal structure
with lattice parameters $a=4.6122(4)$\,{\AA} and
$c=7.5483(9)$\,{\AA}. Microprobe analysis revealed a uniform
distribution of the Cd throughout the sample and an actual
concentration of only about one percent Cd in the sample, i.e.
$\sim10\%$ of the nominal concentration in the flux. For an easier
comparison with literature, we will refer in this paper, too, to the
nominal concentration as used in Ref.~\cite{Pham06}. The magnetic
order was investigated by elastic neutron scattering. The
experiments were performed on the cold triple-axis spectrometer V2
at the BER-II reactor of the Hahn-Meitner-Institut in Berlin, using
a wavelength of the incoming neutrons of $\lambda = 2.73\,{\rm
\AA}$, corresponding to a neutron energy $E=11$ meV. Pyrolytic
graphite, PG(002), was used as monochromator and analyzer. The
horizontal collimation before the monochromator was given by the
$^{58}$Ni guide and was 60' before the sample, before the analyzer
and in front of the detector. A tunable PG-filter in the scattered
beam reduced the contamination of second-order neutrons. The
platelet-like sample of CeCo(In$_{0.9}$Cd$_{0.1}$)$_5$ with a mass
$m\approx 10$ mg and dimensions $2 \times 2 \times 0.3{\rm ~mm}^3$,
0.3 mm being the thickness along the tetragonal $c$-axis, was
mounted on a copper pin attached to the mixing chamber of a dilution
refrigerator. Data were recorded at temperatures between $T=60$~mK
and 3.2~K along principal and high symmetry directions in the ({\em
h}\,{\em h}\,$\ell$\,) scattering plane. Analyzing the scattered
neutrons, i.e. performing elastic scattering, considerably improves
the signal-to-background ratio in comparison to (standard)
diffraction. This holds especially true in our case since the sample
and thus the signal was quite small. In addition to these neutron
scattering studies, we conducted heat capacity and electrical
resisitivity measurements in a Physical Properties Measurement
System (Quantum Design) in the temperature range $350 {\rm ~mK}\leq
T\leq10{\rm ~K}$. In order to correlate the results of the
microscopic and the macroscopic studies, all experiments were
performed on the same single-crystalline sample.

\begin{figure}[t]
\centering\label{Rocking-Scan}
\includegraphics[angle=0,width=75mm,clip]{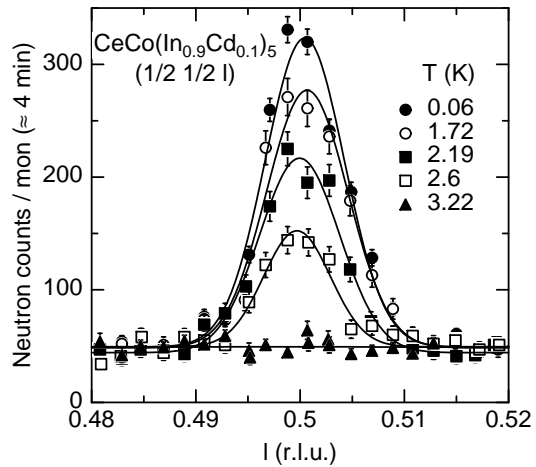}
\caption{Elastic scans along $[001]$ across
$Q=(\frac{1}{2},\frac{1}{2},\frac{1}{2})$ in
CeCo(In$_{0.9}$Cd$_{0.1}$)$_5$ at different temperatures. No
magnetic intensity can be resolved at $T=3.22{\rm ~K}$ ($>T_{\rm
N}$, as determined by specific heat). The solid lines are Gaussian
fits to the data.\label{Rocking-Scan}}
\end{figure}

The specific heat data taken on this sample (see Fig. \ref{Cp_rho})
show two anomalies, one at $T_{\rm N}=3.02$ K corresponding to the
transition to the antiferromagnetically ordered state and the second
one at $T_c=1.27$ K indicating the phase transition to the
superconducting state, in good agreement with
literature.\cite{Pham06} Despite this $T_c$ value, zero resistance
is already observed below 2 K. Similar deviations between
thermodynamic and electrical transport results are also known for
CeIrIn$_5$\cite{Petrovic01a} as well as for Ir-rich
CeRh$_{1-x}$Ir$_x$In$_5$.\cite{Pagliuso01} The specific heat
exhibits a mean-field like jump $\Delta C= 1.46$ J/mol K at $T_{\rm
N}$. Below the AF transition only 30\% of the magnetic entropy ($R
\ln2$) is released suggesting a substantially Kondo-compensated
ordered moment.

To determine the magnetic structure of
CeCo(In$_{0.9}$Cd$_{0.1}$)$_5$ we carried out elastic neutron
scattering experiments and performed elastic scans along
$(\frac{1}{2},\frac{1}{2},\ell)$ at 60 mK. Magnetic intensity was
detected at $Q=(\frac{1}{2},\frac{1}{2},\frac{1}{2})$, cf.
Fig.~\ref{Rocking-Scan}, and at symmetry equivalent positions like
$Q=(\frac{3}{2},\frac{3}{2},\frac{1}{2})$ and
$Q=(\frac{3}{2},\frac{3}{2},\frac{3}{2})$. Due to the small sample
size long counting times (several minutes per point) were required.
The magnetic peak at $(\frac{1}{2},\frac{1}{2},\frac{1}{2})$
monotonically decreases upon heating the sample and vanishes at
$T_{\rm N}$. Scans along other high symmetry directions revealed no
additional intensity, e.g. no intensity was found at
$(1,1,\frac{1}{2})$, $(0,0,\frac{1}{2})$,
$(\frac{3}{2},\frac{3}{2},0)$, or $(\frac{3}{2},\frac{3}{2},1)$. In
particular, no magnetic intensity could be detected at
$(\frac{1}{2},\frac{1}{2},\delta)$, $\delta\approx 0.3$, an
incommensurate position where CeRhIn$_5$, a closely related member
of the Ce$T$In$_5$ ($T={\rm Co}$, Rh, or Ir) family, displays
magnetic superstructure peaks.\cite{Bao00} Hence, the commensurate
magnetic order in CeCo(In$_{0.9}$Cd$_{0.1}$)$_5$ with a propagation
vector $Q_{\rm AF}=(\frac{1}{2},\frac{1}{2},\frac{1}{2})$ is in
marked contrast to the incommensurate order observed in other
compounds of the Ce{\em T}In$_5$ family as well as in
CeCu$_2$Si$_2$.\cite{Stockert04} It is speculated that rather small
changes in the Fermi surface are responsible for this behavior.

The integrated magnetic intensity obtained from Gaussian fits to the
data (cf. Fig. \ref{Rocking-Scan}) is depicted in Fig.
\ref{Intensity}. The magnetic intensity starts to build up below
$T_{\rm N}\approx3$~K, in agreement with $T_{\rm N}$ determined from
specific heat, increases continuously and eventually saturates below
$\approx T_c~ (=1.27$~K), potentially indicating missing magnetic
intensity in the superconducting state. No distinct anomaly is
resolved at the superconducting transition, in particular, the
magnetic intensity does not vanish below $T_c$. Our neutron
scattering results clearly demonstrate the existence of AF order
well below $T_c$, down to lowest temperatures ($T=60$ mK). Nuclear
magnetic quadrupole resonance (NQR) experiments give supplementary
evidence for the microscopic coexistence of SC and
antiferromagnetism, and the estimated moment of approximately
$0.7\mu_{\rm B}$ is in line with the observed magnetic
intensity,\cite{Urbano07} however, we cannot calculate the magnetic
moment precisely.

\begin{figure}[t]
\centering\label{Intensity}
\includegraphics[angle=0,width=75mm,clip]{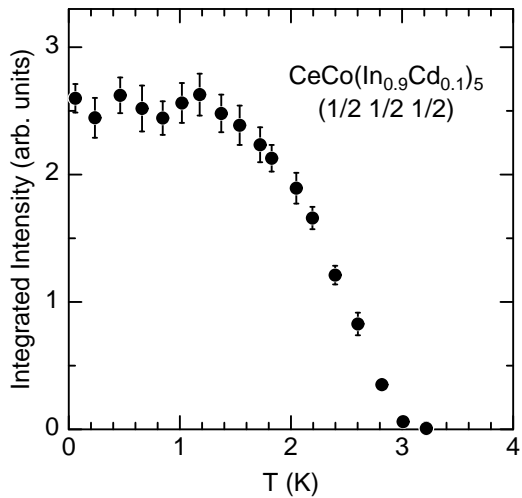}
\caption{Temperature dependence of the integrated magnetic intensity
as obtained by Gaussian fits to the scans across
$Q=(\frac{1}{2},\frac{1}{2},\frac{1}{2})$.\label{Intensity}}
\end{figure}

As shown by Pham et al.,\cite{Pham06} applying pressure reverses the
effect of Cd doping. A generalized pressure-temperature ($p-T$)
phase diagram for CeRhIn$_5$ and the doping series
CeCo(In$_{1-x}$Cd$_x$)$_5$ (including pure
CeCoIn$_5$)\cite{Nicklas01,Sidorov02} describing the pressure
dependence of $T_{\rm N}(p)$ and $T_c(p)$ suggests the same
underlying physics. According to this $p-T$ phase diagram,
CeCo(In$_{0.9}$Cd$_{0.1}$)$_5$ can be considered to correspond to
CeRhIn$_5$ at $p=1.6$ GPa. As already mentioned, at atmospheric
pressure CeRhIn$_5$ orders magnetically below $T_{\rm N}=3.8$ K with
an incommensurate ordering wave vector
$(\frac{1}{2},\frac{1}{2},\delta)$, $\delta=0.297$. Pressure
dependent neutron scattering experiments on CeRhIn$_5$ at 1.8 K show
that the incommensurability, $\delta$, and the ordered moment are
changing only slightly.\cite{Llobet04} In marked contrast to the
simple expectation inferred from our present results, {\it no}
magnetic intensity at 1.6 GPa is reported at
$(\frac{1}{2},\frac{1}{2},\frac{1}{2})$ in CeRhIn$_5$ corresponding
to CeCo(In$_{0.9}$Cd$_{0.1}$)$_5$ at ambient pressure.

In CeRhIn$_5$ AF order can be suppressed by either Co or Ir
substitution on the Rh-site. With increasing substitution level a
superconducting phase, first coexisting with antiferromagnetism,
develops until antiferromagnetism becomes suppressed and only SC
survives.\cite{Pagliuso01,Zapf01} The corresponding phase diagram is
similar to the one of CeCo(In$_{1-x}$Cd$_x$)$_5$ with CeCoIn$_5$
being situated on the purely SC side, cf. Fig.~\ref{PhD}. In Rh-rich
CeRh$_{1-y}$Ir$_y$In$_5$ pressure studies even reveal the same
generic $p-T$ phase diagram found for
CeCo(In$_{1-x}$Cd$_x$)$_5$\cite{Pham06} suggesting a close
relationship.\cite{Nicklas04} In striking contrast to the neutron
scattering results obtained for CeRhIn$_5$ under pressure, a
commensurate magnetic structure with ordering wave vector
$(\frac{1}{2},\frac{1}{2},\frac{1}{2})$ develops in the doping
series CeRh$_{1-y}$Ir$_y$In$_5$ and CeRh$_{1-z}$Co$_z$In$_5$ at low
temperatures, while the same incommensurate ordering wave vector
present in the pure system is still observed below $T_{\rm
N}$.\cite{Christianson05,Yokoyama06} We speculate that the
appearance of the commensurate magnetic ordering wave vector
$(\frac{1}{2},\frac{1}{2},\frac{1}{2})$ is related to SC. Perhaps
this commensurate magnetic ordering has so far been overlooked in
superconducting CeRhIn$_5$ under pressure. Improved experiments with
a better signal-to-background ratio are called for to answer this
question.

In summary, we carried out elastic neutron scattering experiments on
CeCo(In$_{0.9}$Cd$_{0.1}$)$_5$. At low temperatures we found
magnetic intensity at the commensurate wave vector $Q_{\rm
AF}=(\frac{1}{2},\frac{1}{2},\frac{1}{2})$. The magnetic intensity
is building up below $T_{\rm N}$, with $T_{\rm N}$ being in good
agreement with specific heat data. No indication for additional
intensity was observed at incommensurate positions where CeRhIn$_5$,
the related AF member in the Ce$T$In$_5$ family, orders. At low
temperatures magnetic order is coexisting with SC. A saturation of
the magnetic intensity below $T_c$ possibly reveals missing magnetic
intensity in the superconducting state.

We acknowledge useful discussions with P. Thalmeier and G. Knebel.
We would like to thank A. D. Bianchi for the X-ray diffraction and
M. Meissner for his assistance with the low temperature equipment at
HMI. Work at Los Alamos was performed under the auspices of the U.S.
DOE/Office of Science. Work at UC Davis and UC Irvine has been
supported by NSF Grant No. DMR 053360.

\newpage


\begin{thebibliography}{100}

\bibitem{Steglich79} F. Steglich, J. Aarts, C. D. Bredl, W. Lieke,
D. Meschede, W. Franz, and H. Sch\"{a}fer, Phys. Rev. Lett.
  \textbf{43}, 1892 (1979).

\bibitem{Petrovic01b} C. Petrovic, P. G. Pagliuso, M. F. Hundley, R. Movshovich,
J. L. Sarrao, J. D. Thompson, Z. Fisk, and P. Monthoux, J. Phys.
Cond. Mat. {\bf13}, L337 (2001).

\bibitem{Petrovic01a} C. Petrovic, R. Movshovich, M. Jaime, P. G. Pagliuso, M. F. Hundley,
J. L. Sarrao, Z. Fisk, and J. D. Thompson, Europhys. Lett. {\bf53},
354 (2001).

\bibitem{Hegger00}H. Hegger, C. Petrovic, E. G. Moshopoulou, M. F. Hundley, J. L. Sarrao, Z. Fisk,
and J. D. Thompson, Phys. Rev. Lett. {\bf 84}, 4986 (2000).

\bibitem{Kohori01} Y. Kohori, Y. Yamato, Y. Iwamoto, T. Kohara, E. D.
Bauer, M. B. Maple, and J. L. Sarrao, Phys. Rev. B {\bf 64}, 134526
(2001).

\bibitem{Bianchi03}A. Bianchi, R. Movshovich, I. Vekhter, P.G. Pagliuso, and J. L.
Sarrao, Phys. Rev. Lett. {\bf91}, 257001 (2003).

\bibitem{Paglione03}J. Paglione, M. A. Tanatar, D.G. Hawthorn, E. Boaknin, R.W. Hill, F. Ronning,
M. Sutherland, and L. Taillefer, Phys. Rev. Lett. {\bf91}, 246405
(2003).

\bibitem{Pham06} L. D. Pham, Tuson Park, S. Maquilon, J. D. Thompson, and Z.
Fisk, Phys. Rev. Lett. {\bf97}, 056404 (2006).

\bibitem{Mitrovic06}V. F. Mitrovi\'{c}, M. Horvati\'{c}, C. Berthier, G. Knebel, G. Lapertot,
and J. Flouquet, Phys. Rev. Lett. {\bf97}, 117002 (2006).

\bibitem{Young07}B.-L. Young, R. R. Urbano, N. J. Curro, J. D. Thompson, J. L. Sarrao,
A. B. Vorontsov, and M. J. Graf, Phys. Rev. Lett. {\bf98}, 036402
(2007).

\bibitem{Hundley} M. F. Hundley, private communication.

\bibitem{Pagliuso01} P. G. Pagliuso, C. Petrovic, R. Movshovich, D. Hall, M. F.
Hundley, J. L. Sarrao, J. D. Thompson, and Z. Fisk, Phys. Rev. B
{\bf 64}, 100503(R) (2001).

\bibitem{Bao00}W. Bao, P. G. Pagliuso, J. L. Sarrao, J. D. Thompson, Z.
Fisk, J. W. Lynn, and R. W. Erwin,  Phys. Rev. B {\bf62}, R14621
 (2000); W. Bao, P. G. Pagliuso, J. L. Sarrao, J. D. Thompson, Z.
Fisk, J. W. Lynn, and R. W. Erwin,  Phys. Rev. B {\bf63}, 219901(E)
(2001).

\bibitem{Stockert04}O. Stockert, E. Faulhaber, G. Zwicknagl, N. St\"{u}{\ss}er, H. S. Jeevan,
M. Deppe, R. Borth, R. K\"{u}chler, M. Loewenhaupt, C. Geibel, and
F. Steglich, Phys. Rev. Lett. {\bf 92}, 136401 (2004).

\bibitem{Urbano07} R. R. Urbano, B.-L. Young, N. J. Curro, J. D. Thompson, L. D. Pham, and Z.
Fisk, cond-mat/0702008 (2007).

\bibitem{Nicklas01}M. Nicklas, R. Borth, E. Lengyel, P. G. Pagliuso, J. L. Sarrao,
V. A. Sidorov, G. Sparn, F. Steglich, and J. D. Thompson1, J. Phys.
Condens. Matter {\bf13}, L905 (2001).

\bibitem{Sidorov02} V. A. Sidorov, M. Nicklas, P.G. Pagliuso, J. L. Sarrao, Y.
Bang, A.V. Balatsky, and J. D. Thompson, Phys. Rev. Lett. {\bf89},
157004 (2002).

\bibitem{Llobet04}A. Llobet, J. S. Gardner, E. G. Moshopoulou, J.-M. Mignot, M. Nicklas, W. Bao, N. O. Moreno,
P. G. Pagliuso, I. N. Goncharenko, J. L. Sarrao, and J. D. Thompson,
Phys. Rev B {\bf69}, 024403 (2004).

\bibitem{Zapf01} V. S. Zapf, E. J. Freeman, E. D. Bauer, J. Petricka, C. Sirvent,
N. A. Frederick, R. P. Dickey, and M. B. Maple, Phys. Rev. B {\bf
65}, 014506 (2001).

\bibitem{Nicklas04} M. Nicklas, V. A. Sidorov, H. A. Borges, P. G. Pagliuso, J. L. Sarrao, and J. D.
Thompson, Phys. Rev. B, {\bf 70}, 020505(R) (2004).

\bibitem{Christianson05} A. D. Christianson, A. Llobet, Wei Bao, J. S. Gardner, I. P.
Swainson, J.W. Lynn, J.-M. Mignot, K. Prokes, P. G. Pagliuso, N. O.
Moreno, J. L. Sarrao, J. D. Thompson, and A. H. Lacerda, Phys. Rev.
Lett. {\bf 95}, 217002 (2005).

\bibitem{Yokoyama06} M. Yokoyama, H. Amitsuka, K. Matsuda, A. Gawase, N. Oyama, I.
Kawasaki, K. Tenya, and H. Yoshizawa, J. Phys. Soc. Jpn. {\bf75},
103703 (2006).

\end{thebibliography}
\end{document}